\documentclass[letterpaper,twocolumn,english,citesort,noshowpacs,noshowkeys,preprintnumbers,prd,floatfix,nofootinbib,superscriptaddress]{revtex4}
\usepackage[T1]{fontenc}
\usepackage[latin1]{inputenc}
\usepackage{amsmath}
\usepackage{color} 
\usepackage{graphicx}
\usepackage{amssymb}

\makeatletter

\providecommand{\tabularnewline}{\\}

\usepackage{babel}
\makeatother

\begin{document}

%\preprint{CERN----}
%\preprint{hep-ph/**********}
%\eprint{hep-ph/**********}

\title{Correlated theoretical uncertainties for the one-jet inclusive cross
section}

\author{Fredrick I. Olness}  
\email{olness@smu.edu}

\affiliation{Department of Physics, Southern Methodist University, Dallas, TX
75275-0175, USA }

\author{Davison E. Soper}
\email{soper@uoregon.edu}

\affiliation{Institute of Theoretical Science, University of Oregon Eugene, OR
97403-5203, USA}

\date{\today}

\begin{abstract}
We discuss the correlated systematic theoretical uncertainties that
may be ascribed to the next-to-leading order QCD theory used to predict
the one-jet inclusive cross section in hadron collisions. We estimate
the magnitude of these errors as functions of
the jet transverse momentum and rapidity. The total theoretical
error is decomposed into a set of functions of transverse momentum and rapidity that give a model for statistically independent contributions to the error. This representation can be used to include the systematic theoretical errors in fits to the experimental
data. 
\end{abstract}
\maketitle
\tableofcontents{}

\newpage{}

\section{Introduction\label{sec:Introduction}}

Predictions of the Standard Model are typically made with the aid
of next-to-leading order (NLO) perturbative calculations (or sometimes
with NNLO calculations). Evidently, these predictions are not exactly
equal to what one should measure if the Standard Model is correct.
If we have an NLO calculation, we leave out NNLO and ${\rm N}^{3}{\rm LO}$
contributions, \emph{etc}. We also leave out contributions that are
suppressed by a power of the large momentum scale of the problem.
Of course, we do not know exactly how big these contributions are:
if we could calculate them, we would include them in the prediction.
Nevertheless, we can estimate the size of the corrections. They then
constitute ``theory errors'' in the prediction, which are quite
similar to experimental systematic errors in the measurement.

In this paper we distinguish between errors associated with higher
order contributions and power suppressed contributions to the cross
section, which we call theory errors, and errors associated with our
imperfect knowledge of the parton distribution functions needed for
the prediction. Estimated theory errors are needed in two contexts.
First, if an experiment does not agree with the theoretical prediction
within the experimental statistical and systematic errors, then we
need to see if there is agreement within the combined experimental
and theory errors and the errors from the parton distributions used
in the prediction. In the case that the disagreement is outside of
the combined errors, then we have a signal for new physics.

The second context in which we need estimated theory errors is in
the determination of parton distribution functions from experimental
measurements. The theory errors give a contribution to the errors
that we associate with the parton distribution functions that emerge
from a fit to the data. Evidently, if we do not include theory errors,
the resulting errors in the parton distribution functions will be
too small. Additionally, if for one kind of process the theory errors
are large while for another kind of process the theory errors are small,
then we will give the large-error process too much weight in the fit.

In this paper, we provide an estimate of the theory error for the
one jet inclusive cross section $d^{2}\sigma/dP_{T}\, dy$ in hadron-hadron
collisions, where $P_{T}$ is the transverse momentum or {}``transverse
energy'' of the jet, and $y$ is the rapidity of the jet. There is good data
for this process from the CDF and D0 experiments at Fermilab, including
careful estimates of the experimental systematic errors. Estimates
of the theory errors are needed to accompany the estimates of the experimental
systematic errors.

We warn that there is no unique method to estimate theory errors.
Thus our task is to provide a method that is defensible if not necessarily
optimal. We seek to provide an estimate in a form that includes the
correlations from one $\{P_{T},y\}$ point to another.

\section{General Setup\label{sec:General-setup}}

\begin{figure*}
\begin{centering}
\includegraphics[width=0.45\textwidth]{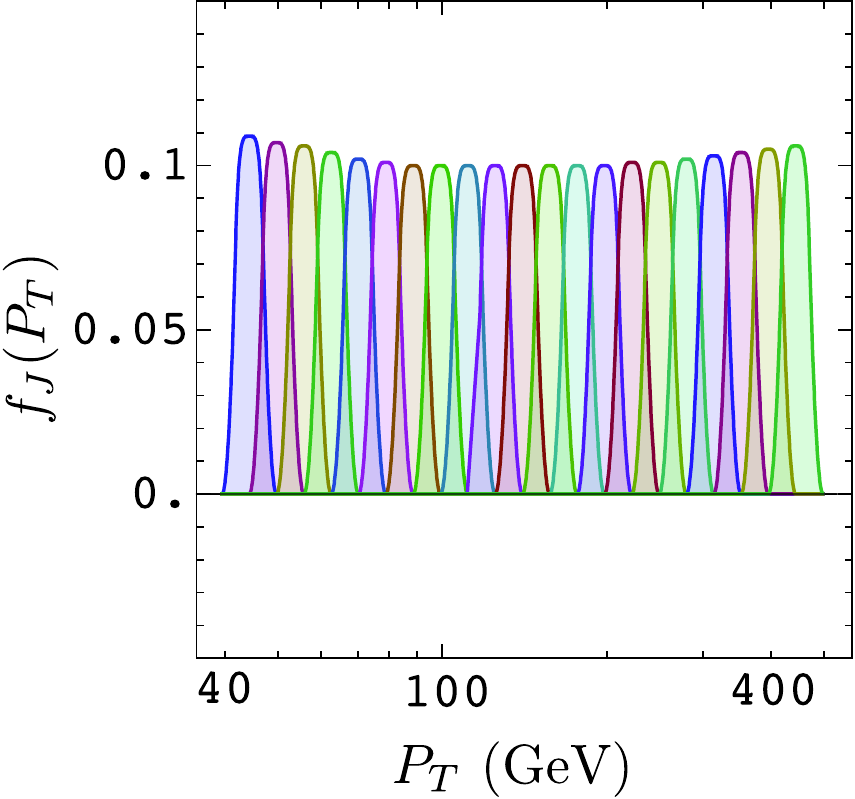}\qquad{}\includegraphics[width=0.45\textwidth]{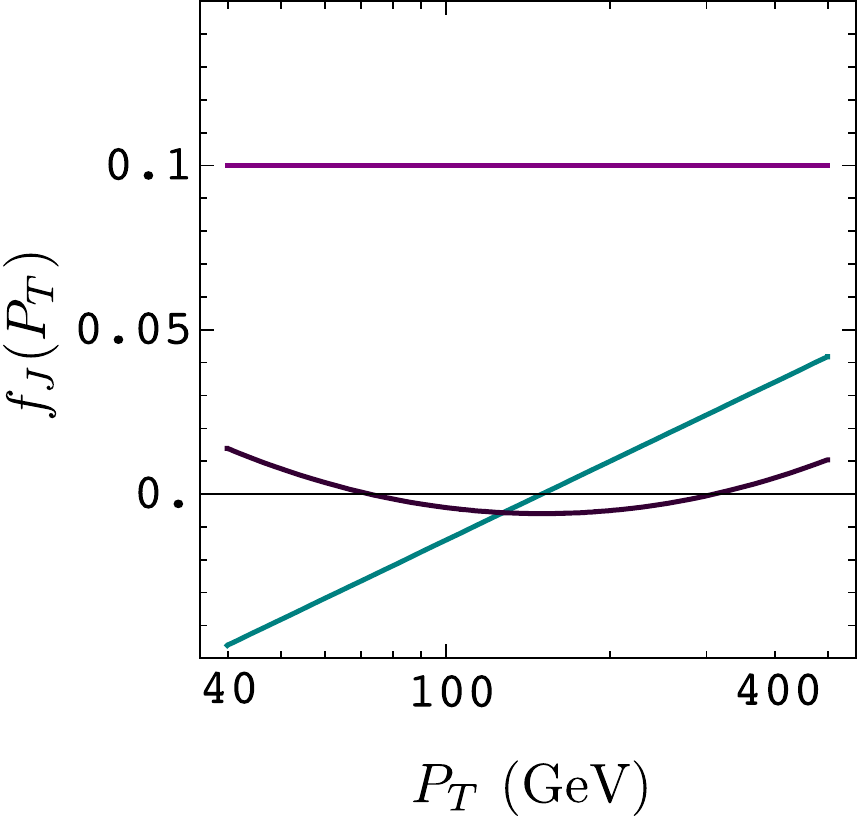}
\par\end{centering}

\caption{illustration of (a) uncorrelated and (b) correlated theoretical errors.
In (a), the total error is about 10\% for all $P_{T}$, but the error
at any $P_{T}$ is not correlated with the error at nearby points.
In (b), there are just three functions $f_{J}(P_{T})$ giving, again,
about a 10\% total error at any one $P_{T}$. Because the $f_{J}(P_{T})$
are smooth functions, the theoretical error at a given $P_{T}$ will
be smoothly related to the error at other $P_{T}$ values.\label{fig:correlation}}

\end{figure*}

We treat theory errors in a fashion that is similar to that used for correlated systematic errors in the experimental results. 
We use next-to-leading order quantum chromodynamics (QCD)  theory to make predictions
for the one-jet inclusive cross 
section\footnote{Specifically, we use the program of Ref.~\citep{Ellis:1990ek},
although there are other programs that can give the same results.
The code is available at http://zebu.uoregon.edu/$\sim$soper/EKSJets/jet.html }
\begin{eqnarray*}
\frac{d\sigma}{dP_{T}\, dy}&& 
\\ && \hskip - 1 cm =  \int\! dx_{1}\int\! dx_{2}\, 
f_{a/A}(x_{1},\mu)\, f_{b/B}(x_{2},\mu)\,
\frac{d\hat{\sigma}_{ab\to {\rm jet}}}{dP_{T}\, dy}
\;\;.
\end{eqnarray*}
In the calculation, one uses Monte Carlo integration so that there
is a random statistical error for each point $\{P_{T},y\}$. We do
not include these statistical errors in the analysis here since they
are typically quite small (say 2\%) and one can reduce them by running
the program for a longer time. If we wished to include the errors
from fluctuations in the Monte Carlo integrations, that task would
be straightforward because the statistical nature of these fluctuations
is known.

We will start our investigation by studying jet production
corresponding to the Tevatron Run 2, with $\sqrt{s} = 1960\ {\rm
GeV}$, as a function of $P_T$ and $y$. We will display the results for
$y = \{0,1,2\}$ as functions of $P_T$; we also present formulas for
the $P_T$ and $y$ dependence, from which estimated errors for the
specific kinematic ranges used by CDF and D0 can be inferred.

We need estimated errors that can be used in a statistical analysis. However, we do not have at hand a statistical ensemble of worlds in which terms beyond those included in the NLO theory vary.  Thus we make estimates that we hope are reasonable but that can and should be subject to debate. 

We formulate the treatment of theory errors as follows. We let
\begin{equation}
\frac{d\sigma}{dP_{T}dy}=\left[\frac{d\sigma}{dP_{T}dy}\right]_{{\rm NLO}}\left\{ 1+\sum_{J}\lambda_{J}f_{J}(P_{T},y)\right\} 
\;\;.\label{eq:errorstructure}
\end{equation}
Here the functions $f_{J}(P_{T},y)$ are definite functions, while
the $\lambda_{J}$ are unknown parameters. Thus $\lambda_{J}f_{J}(P_{T},y)$
represents an unknown theoretical contribution that might modify the
NLO theory. We treat the $\lambda_{J}$ as Gaussian random variables
with variance 1. That is, the size of the uncertainty with label $J$
is represented by how big $f_{J}(P_{T},y)$ is. 
If one thinks of this as representing an imaginary ensemble of worlds in which theory calculations come out differently, then these worlds all have the same $f_J$ but the $\lambda_J$ vary.

We will propose to use just a few functions $f_J$. We offer 
the following defense of this strategy. Consider a simplified case of a 
cross section that is a function of just one variable, $P_T$. If we were
to believe that the uncertainty in the prediction of this cross section is of order, say, 10\%, but we have
no idea of what the shape of the true cross section is within a 10\%
band about the prediction, then we would choose many functions
$f_{J}(P_{T})$, each of size 0.10, but with each being non-zero
only in a very tiny range of $P_{T}$. This approach is illustrated
in Figure~\ref{fig:correlation}-a); such a view seems to us unreasonable. 

Experience with various perturbative and non-perturbative contributions
teaches that they are smooth functions of the relevant variables,
$P_{T}$ in this case. This arguably more reasonable scenario is illustrated
in Figure~\ref{fig:correlation}-b). 
As illustrated by the three
curves,%
\footnote{Specifically, in this figure we use the functions $f_{1}(P_{T})=0.1$,
$f_{2}(P_{T})=0.08\,\log(P_{T}/M)$, and $f_{3}(P_{T})=0.06\,\left\{ \left[\log(P_{T}/M)\right]^{2}-0.1\right\} $
where $M$= 150~GeV. These curves are for illustrative purposes only,
and the $f_{J}(P_{T})$ functions differ from the set $f_{J}(P_{T},y)$
we will use to parameterize the correlated systematic uncertainties.%
} one contribution beyond NLO could be flat, amounting to a constant ``K factor,'' another might be a smoothly increasing function of $P_T$, while yet another might be positive at high and low $P_T$ and negative in between. However, we judge it unlikely that a currently uncalculated contribution contribution would have multiple maxima between low and high $P_T$.

Thus we seek a few functions $f_{J}(P_{T},y)$ that have some dependence
on $\{P_{T},y\}$ and represent, as best we can determine, our understanding
of the character of uncalculated contributions. In the following sections,
we analyze several sources of theory errors and associate them with
functions $f_{J}(P_{T},y)$.

\section{Perturbative uncertainty\label{sec:ScaleDependence}}

The main source of uncertainty at large jet transverse momentum, at least
in our estimation, is the fact that we have calculated only at NLO, leaving contributions from higher orders of perturbation uncalculated. We estimate this uncertainty using the dependence of the computed cross section on the renormalization and factorization scales. We present this estimate in this section. In the following section, we check this estimate using an independent method involving threshold effects.

\subsection{Error estimate from scale dependence}

The first ingredient in our estimation of theory errors is based on
the traditional method in which one evaluates the dependence of the
computed NLO cross section on two scales: the renormalization scale
$\mu_{{\rm R}}$ and the factorization scale $\mu_{{\rm F}}$.
One often makes a standard choice for these
scales: $\mu_{{\rm R}}=\mu_{{\rm F}}=P_{T}/2$. We will take this
choice as our central value and define \begin{equation}
\begin{split}
x_{1}= & \log_{2}\left(\frac{\mu_{{\rm R}}}{P_{T}/2}\right)\;\;,
\\
x_{2}= & \log_{2}\left(\frac{\mu_{{\rm F}}}{P_{T}/2}\right)\;\;.
\end{split}
\end{equation}
We compute the cross section near $x_{1}=x_{2}=0$, that is near the
scale choice $\mu_{{\rm R}}=\mu_{{\rm F}}=P_{T}/2$. Then $\{x_{1},x_{2}\}$
measures (logarithmically) the distance from this central value. We
then fit the cross section to a quadratic polynomial $P(\vec{x})$
in $\vec{x}$-space,
\begin{equation}
\left[\frac{d\sigma(x_{1},x_{2})}{dP_{T}}\right]_{{\rm NLO}}
\approx
\left[\frac{d\sigma(0,0)}{dP_{T}}\right]_{{\rm NLO}}
\big[1 + P(\vec{x})\big]
\;\;,
\label{eq:Pdef}
\end{equation}
where
\begin{equation}
P(\vec{x}) = \sum_{J}x_{J}A_{J}+\sum_{J,K}x_{J}M_{JK}x_{K}
\;\;,
\label{eq:Pform}
\end{equation}
with $\vec{x}=(x_{1},x_{2})$ and $J,K=\{1,2\}$. 

We know that if we had an NNLO calculation, the dependence of the cross
section on $\vec{x}$ would be canceled to order $\alpha_{s}^{2}$.
Thus the coefficients $A_{J}$ and $M_{JK}$ carry information about
the perturbative coefficients beyond NLO. For this reason, we use
the coefficients $A_{J}$ and $M_{JK}$ to provide an estimate of
the error induced by truncating the perturbative expansion at one-loop
order. We define a simple recipe for this purpose. We define an estimated
error%
\footnote{We shall use ${\cal E}_{\rm scale}$ to denote the theoretical systematic
error due to scale dependence only, and ${\cal E}$ (no subscript)
to denote the total theoretical systematic error.
}%
\ ${\cal E}_{{\rm scale}}$ as the root-mean-square average of
$P(\vec{x})$ over a circle with a certain radius $|\vec{x}|$, \begin{equation}
{\cal E}_{{\rm scale}}^{2}=\frac{1}{2\pi}\int_{0}^{2\pi}\! d\theta\ P(|\vec{x}|\cos\theta,|\vec{x}|\sin\theta)^{2}\;\;.\label{eq:Escale}\end{equation}
 We need to select a value of $|\vec{x}|$, and we make the choice
\begin{equation}
|\vec{x}|=2\;\;.\label{eq:absxchoice}\end{equation}
 In the most common method of estimating errors from scale variation,
we would vary $(2\mu_{{\rm R}}/P_{T},2\mu_{{\rm F}}/P_{T})$ between
$(1,1)$ and $(2,2)$ and between $(1,1)$ and $(1/2,1/2)$. This
amounts to changing $\vec{x}$ from $0$ to a vector of length $|\vec{x}|=\sqrt{2}$
in a particular direction that corresponds to something close to the
direction of strongest variation. The choice $|\vec{x}|=2$ is somewhat
larger than this standard choice. For instance, $|\vec{x}|=2$ in
the direction $\vec{x}\propto(1,1)$ corresponds to \begin{equation}
\left(\frac{\mu_{{\rm R}}}{P_{T}/2},\frac{\mu_{{\rm F}}}{P_{T}/2}\right)=(2^{\sqrt{2}},2^{\sqrt{2}})\approx(2.7,2.7)\;\;.\end{equation}
We average over the directions of $\vec{x}$ instead of taking a
particular direction. For this reason, the value of Eq.~(\ref{eq:absxchoice})
gives results that are similar to the method that is often 
used. 
While varying the $\mu$-scales along the $(1,1)$ direction 
will often work, our averaging technique
provides a general method that seems sensible even when the one of the directions of slowest variation happens to align with the $(1,1)$ direction.

A straightforward calculation shows that, with the definition~(\ref{eq:Escale}),
\begin{equation}
{\cal E}_{{\rm scale}}^{2}=\frac{|\vec{x}|^{2}}{2}\,\vec{A}^{\,2}+\frac{|\vec{x}|^{4}}{8}\left[({\rm Tr}\, M)^{2}+2\,{\rm Tr}\, M^{2}\right]\;\;.\label{eq:Escaleresult}\end{equation}
 We determine the coefficients $A_{J}$ and $M_{JK}$ by calculating
the one jet inclusive cross section for a given value of $P_{T}$
and rapidity. We use nine points in $\vec{x}$-space, obtained by
setting each $\{\mu_{{\rm R}},\mu_{{\rm F}}\}$ scale to $\{\frac{1}{4}P_{T},\,\frac{1}{2}P_{T},\, P_{T}\}$
and fit the results to the form given in Eq.~(\ref{eq:Pdef}) and
Eq.~(\ref{eq:Pform}).

\subsection{Contour plots}

\begin{figure}[!t]
\includegraphics[width=0.45\textwidth]{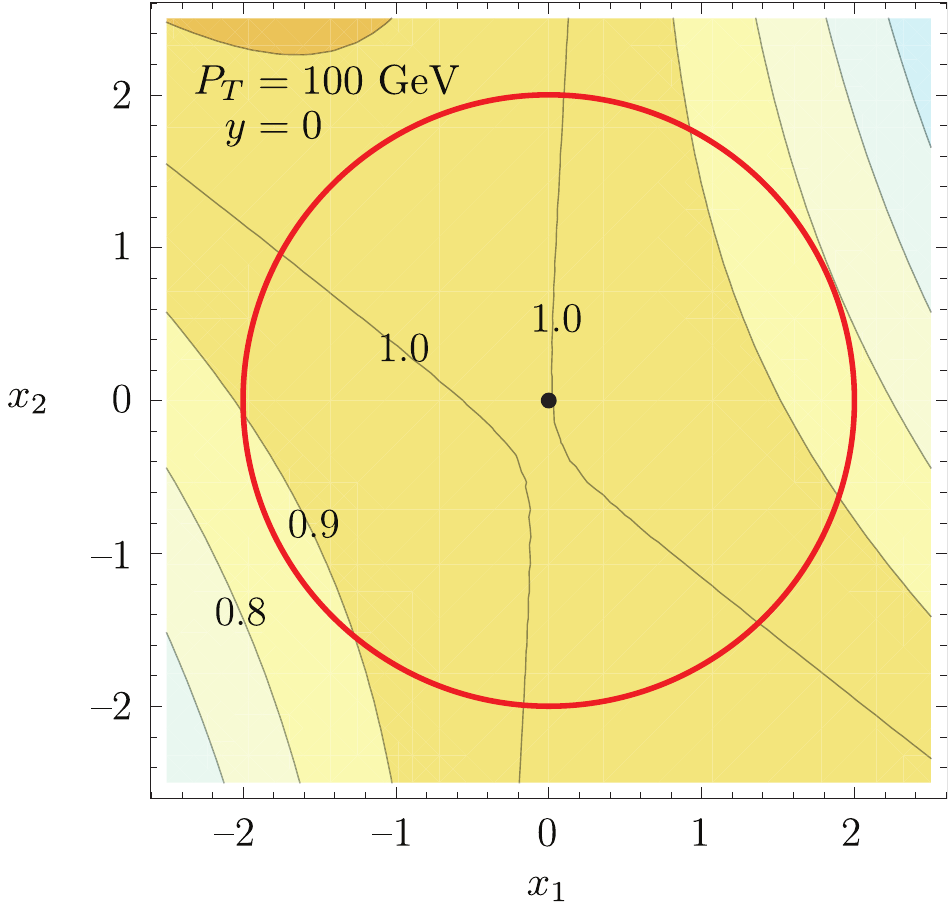}

\includegraphics[width=0.45\textwidth]{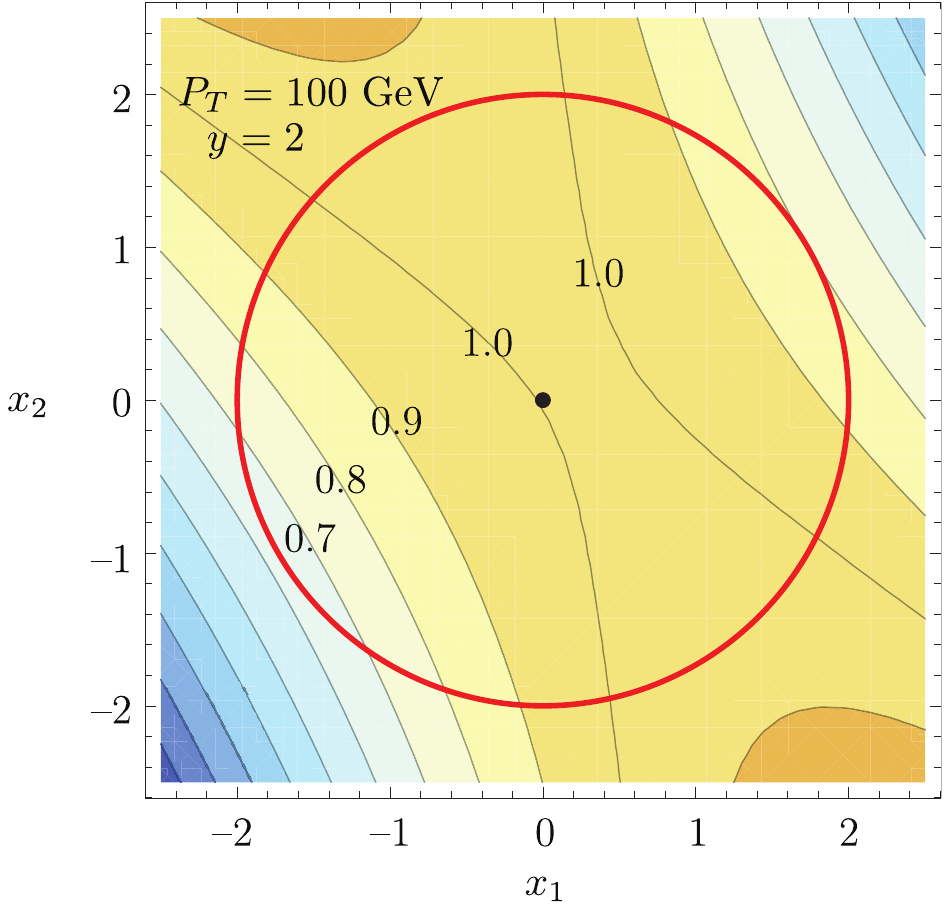}

\caption{Contour plot of the jet cross section in the $\{x_{1},x_{2}\}$ plane
for the Tevatron ($\sqrt{s}=1960$~GeV) with $P_{T}=100$~GeV and
a) central rapidity $y=0$ and b) forward rapidity $y=2$. We plot
the ratio of the cross section compared to the central value at $\{x_{1},x_{2}\}=\{0,0\}$.
Contour lines are drawn at intervals of 0.10. The (red) circle is
at radius $|x|=2$.\label{fig:contours}}

\end{figure}

We illustrate this procedure for estimating the theoretical error
from this source in 
Fig.~\ref{fig:contours}, where we display contour plots of $1 + P(\vec x)$ corresponding to the jet cross section at the Tevatron with $P_{T}=100$~GeV for $y = 0$ and for $y = 2$. For both
values of $y$, we find a saddle point in the vicinity of $\{x_{1},x_{2}\}=\{0,0\}$
which corresponds to $\{\mu_{{\rm R}},\mu_{{\rm F}}\}=\{{P_{T}}/{2},{P_{T}}/{2}\}$.
This location of the saddle point is a general feature that holds
throughout much of the kinematic range; it motivates the choice
$\{\mu_{{\rm R}},\mu_{{\rm F}}\}=\{{P_{T}}/{2},{P_{T}}/{2}\}$
as our central values. 

The estimated scale dependence error, ${\cal E}_{{\rm scale}}$, is
then obtained by averaging the deviation of the cross section at a
given radius in $\vec{x}$-space. As discussed above, we choose a
radius of  $|\vec{x}|=2$, as indicated by the circle in Figure~\ref{fig:contours}.
The slope of the $\{x_{1},x_{2}\}$ surface is steeper for the $y=2$
case as compared with the $y=0$ case. Consequently, we find a larger ${\cal E}_{{\rm scale}}$
for $y=2$ ($\sim18\%$) as compared to $y=0$ ($\sim9\%$).

\subsection{Comment on the range of scale choices}

\begin{figure}[t]
\includegraphics[width=0.48\textwidth,keepaspectratio]{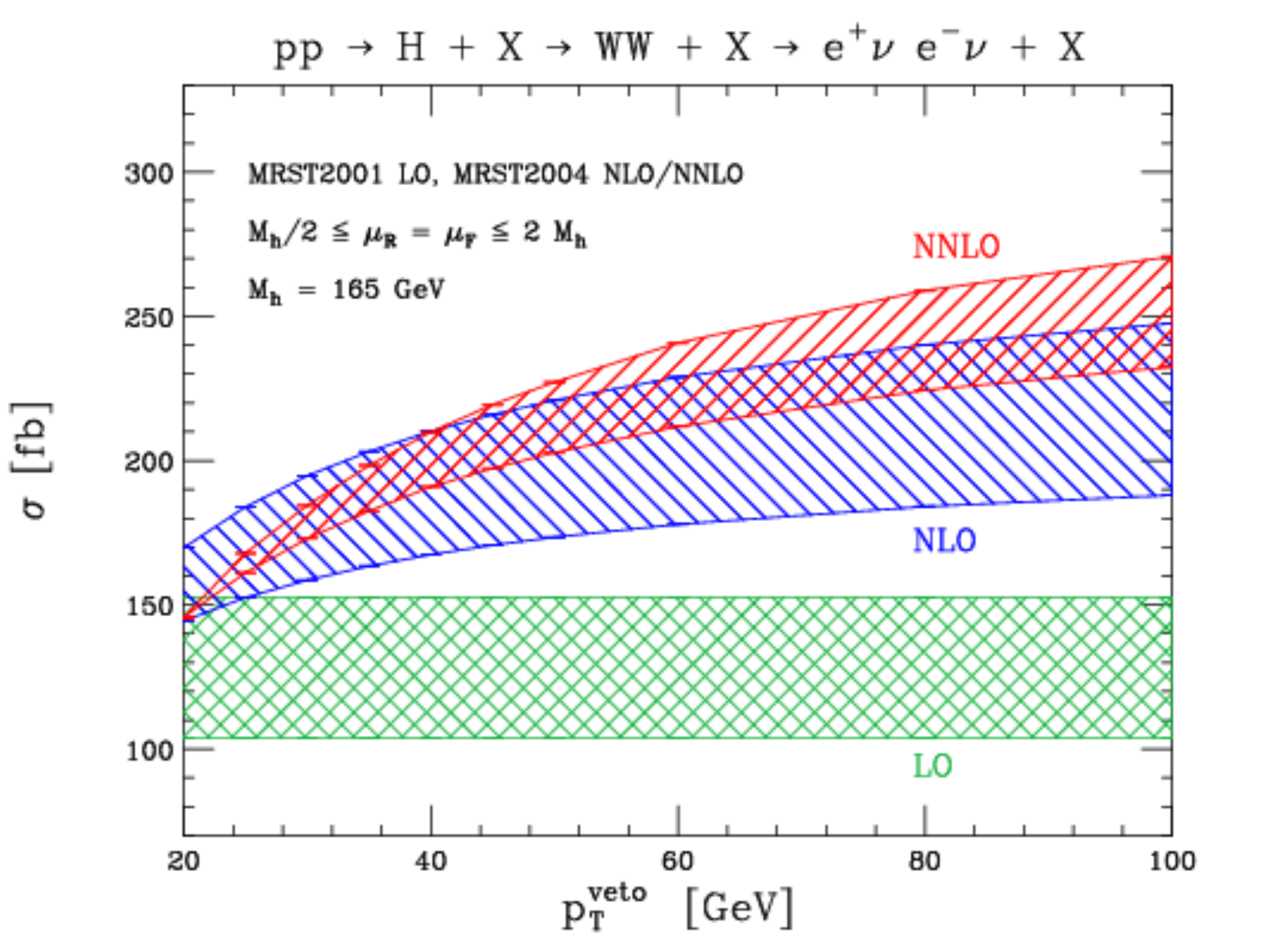}

\caption{The cross section for Higgs production at the LHC for LO, NLO, and
NNLO calculations as taken from Ref.~\citep{Anastasiou:2007mz}. The
computed cross section vetos jets ($P_{T}^{\rm jet}>P_{T}^{\rm veto}$) in
the central region $|\eta|<2.5$. \label{fig:babis}}

\end{figure}

In the above analysis, we estimate the theoretical uncertainty by
varying the $\mu$ scales by a factor about a central value. This
is a conventional choice, but is it reasonable? To examine this question,
one can look at cases in which NNLO calculations exist. Here, we choose
one typical case as an example. In Fig.~\ref{fig:babis}, we show
the NNLO cross section for Higgs production at the Large Hadron Collider (LHC) as a function
of the $P_{T}^{{\rm veto}}$ parameter as calculated by Ref.~\citep{Anastasiou:2007mz}.
Here, the renormalization and factorization scales are varied by a
factor of two, $\{\mu_{\rm R},\mu_{\rm F}\}\in[M_{\rm h}/2,\,2M_{\rm h}]$. 

Consider, for example, $P_T^{\rm veto} $ near 80 GeV. To simplify our
argument, let us suppose that the exact QCD result is known and that
it lies in the middle of the NNLO error band. We then ask whether the
estimated NLO error band was reasonable, now that we know the exact
answer. To do a real statistical analysis, we should have at hand many
NLO calculations of separate and independent quantities, each with its
error estimate. For each such quantity, a NNLO calculation that we can
regard as nearly ``exact'' should be available. We would then plot the
distribution of the differences between the NLO central value and the
true answer in units of the NLO $1\ \sigma$ error estimate. If the
error estimates are reliable, this distribution should be a Gaussian
distribution with width 1. We cannot do that with just one
datum. However, we can say that if the NLO estimate is reasonable then
the central NNLO value in the one case that we have should be roughly
$1\ \sigma$ away from the NLO central value. If it is $3\ \sigma$
away, then it seems likely that the NLO error was underestimated. If
it is $0.1\ \sigma$ away, then seems likely that the NLO error was
overestimated. In the case at hand, the difference is about $1\
\sigma$, so we have some evidence that the error was correctly
estimated.

\subsection{Scale dependence total uncertainty}

\begin{figure*}[t]
\includegraphics[width=0.30\textwidth]{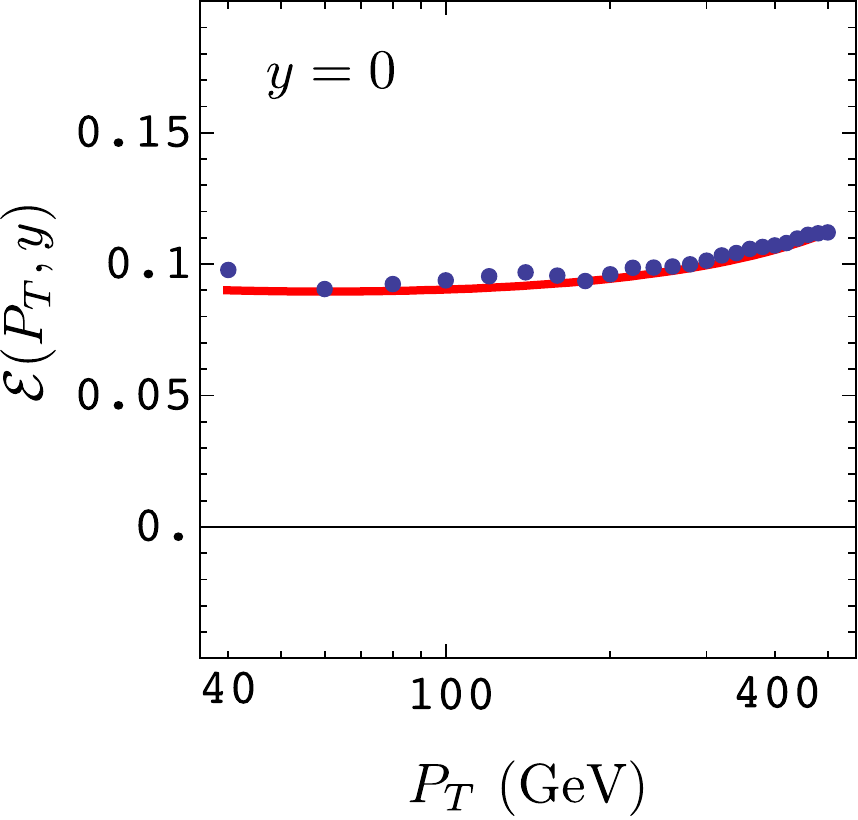}\hskip 0.04\textwidth
\includegraphics[width=0.30\textwidth]{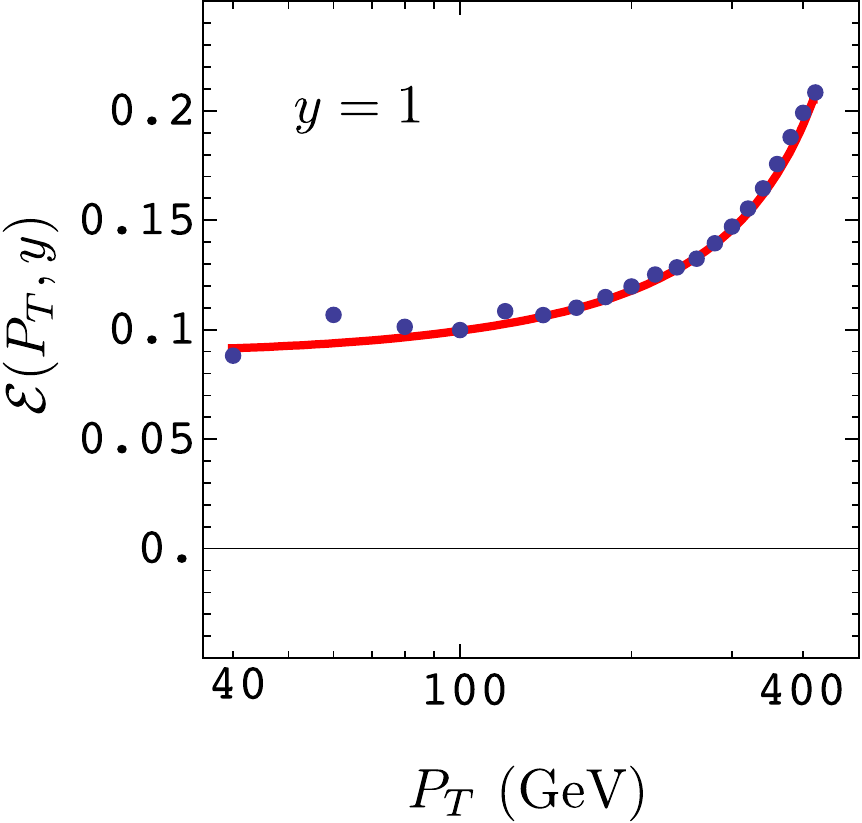}\hskip 0.04\textwidth
\includegraphics[width=0.30\textwidth]{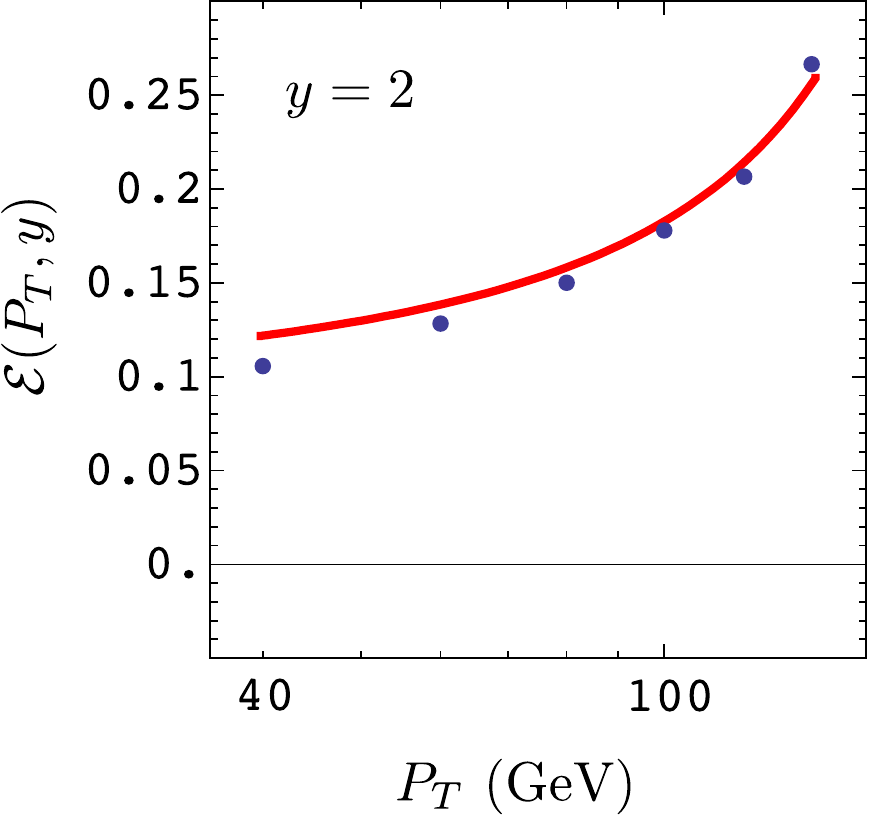}

\caption{The estimate of the uncertainty ${\cal E}(P_T,y) = {\cal E}_{\rm scale}$ due to the scale
variation as given in Eq.~(\ref{eq:Escaleresult}) for the Tevatron
($\sqrt{s}=1960$~GeV) with $y=\{0,1,2\}$. The calculation from
the jet code is represented by the (blue) points, and the fit based
on Eq.~(\ref{eq:neterror}) is shown with the solid (red) curve.
\label{fig:muError0a} }

\end{figure*}

Implementing the procedure outlined above, we find the theoretical
systematic error estimated from scale dependence, 
${\cal E}_{\rm scale}$; this
is displayed in Fig.~\ref{fig:muError0a}.
The (blue) points are ${\cal E}_{\rm scale}$ computed as described above from the NLO cross section \citep{Ellis:1990ek}
and the (red) curve is a smooth fit to these points. 

We see that ${\cal E}_{\rm scale}(P_{T},y)$,
is a slowly rising function of $P_{T}$. For the rapidity $y=0$ at
the Tevatron ($\sqrt{s}=1960\ {\rm GeV}$), we find that ${\cal E}_{\rm scale}(P_{T},y)$
varies from 9\% to 11\%. For
$y=1$, the uncertainty ranges from 9\% to 20\%, and for $y=2$ the
uncertainty increases even more, ranging from 12\% to 25\% over a more
limited $P_{T}$ range.

\subsection{Scale dependence correlated uncertainty}

\begin{figure*}[t]
\includegraphics[width=0.30\textwidth]{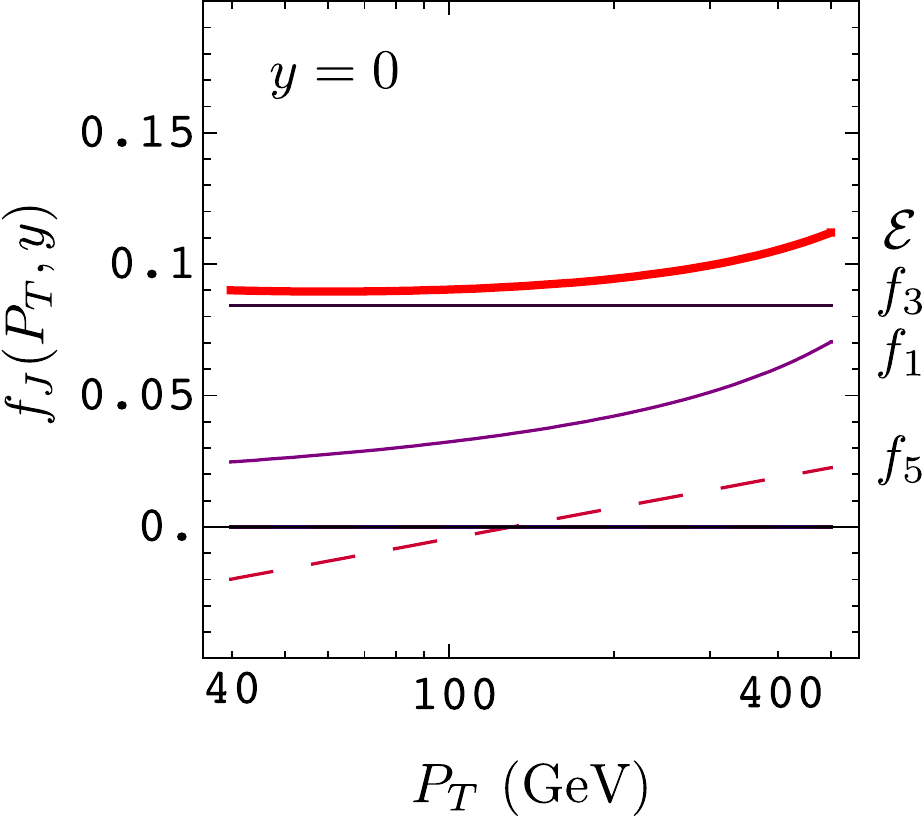} \hskip 0.04\textwidth
\includegraphics[width=0.30\textwidth]{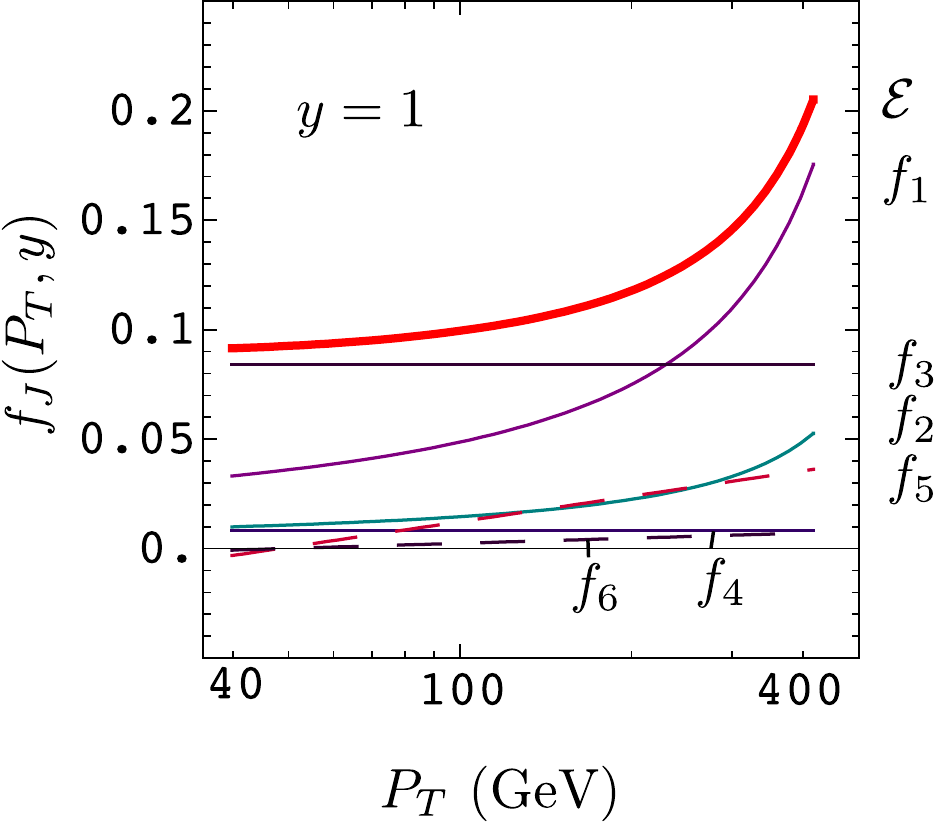} \hskip 0.04\textwidth
\includegraphics[width=0.30\textwidth]{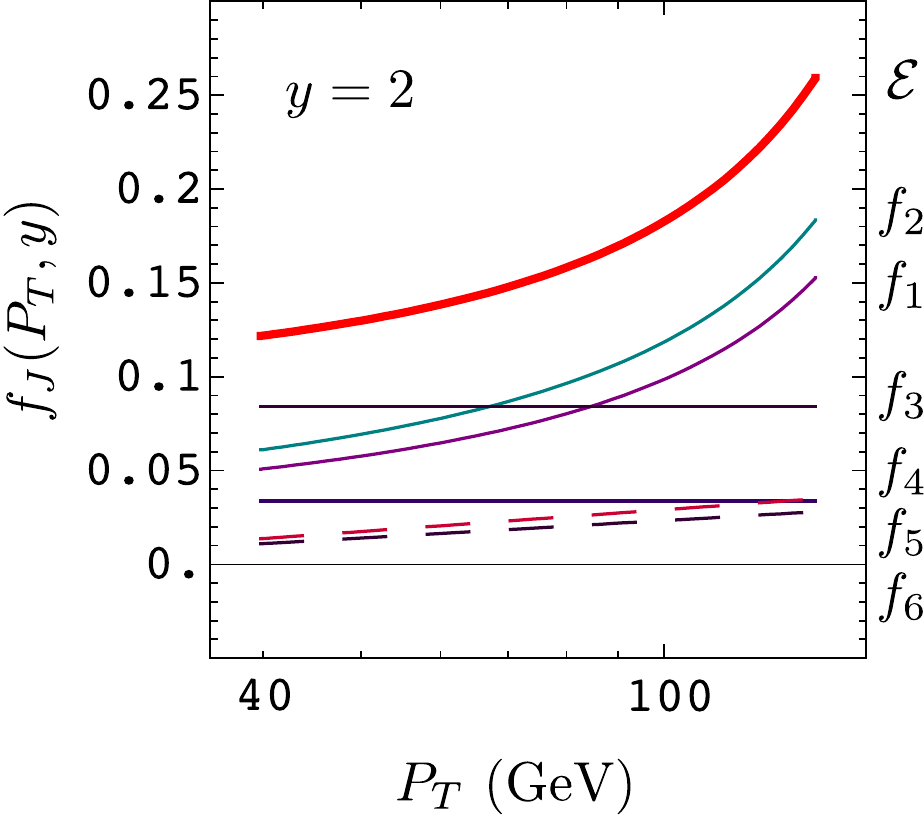}

\caption{The estimate of the uncertainty ${\cal E}_{\rm scale}$ due to the scale
variation as given in Eq.~(\ref{eq:Escaleresult}) for the Tevatron
($\sqrt{s}=1960$~GeV) with $y=\{0,1,2\}$. The combined uncertainty
${\cal E}_{\rm scale}$ is shown as the upper thick (red) curve, and the
individual functions $f_{J}(P_{T},y)$ are indicated below. \label{fig:muError0b} }

\end{figure*}

As described
in Section~\ref{sec:General-setup}, we decompose the total scale dependence 
uncertainty, ${\cal E}_{\rm scale}$,
into a (small) number of functions $f_{J}(P_{T},y)$ which then combine
to form the total uncertainty ${\cal E}_{\rm scale}$. 

Since the $f_{J}(P_{T},y)$ functions represent independent sources of
uncertainty, ${\cal E}_{\rm scale}$ is the quadrature sum 
\begin{equation}
{\cal E}_{\rm scale}(P_{T},y)\equiv\sqrt{\sum f_{J}(P_{T},y)^{2}}
\;\;.\label{eq:neterror}
\end{equation}
We chose a set of functions $f_{J}(P_{T},y)$ that
satisfies Eq.~(\ref{eq:neterror}). We take the $f_{J}(P_{T},y)$ 
to depend on $y$ and on the ratio of $P_{T}$ to the quantity%
\footnote{We scale $P_{T}$ by $M(y)$ to make the argument of the logarithms
dimensionless. This quantity provides a simple scaling, and roughly
corresponds to scaling by the maximum $P_T$, $P_{T}^{\rm max}\sim\sqrt{s}/(2\cosh(y))$, 
for large $y$. %
} \begin{equation}
M(y)=\sqrt{s}\ e^{-y}\;\;.\end{equation}
 For the set of $f_{J}(P_{T},y)$ functions we choose \begin{equation}
\begin{split}f_{1}(P_{T},y)={} & \frac{9.62\times10^{-2}}{\log(M(y)/P_{T})}\;\;,\\
f_{2}(P_{T},y)={} & \frac{2.89\,\times10^{-2}\,\, y^{2}}{\log(M(y)/P_{T})}\;\;,\\
f_{3}(P_{T},y)={} & 8.42\times10^{-2}\;\;,\\
f_{4}(P_{T},y)={} & 0.842\times10^{-2}\, y^{2}\;\;,\\
f_{5}(P_{T},y)={} & 1.68\times10^{-2}\,\log\!\left(\frac{15\, P_{T}}{M(y)}\right)\;\;,\\
f_{6}(P_{T},y)={} & 0.336\times10^{-2}\, y^{2}\log\!\left(\frac{15\, P_{T}}{M(y)}\right)\;\;.\end{split}
\label{eq:errorSummary}\end{equation}
These functions are illustrated in Fig.~\ref{fig:muError0b}.
The first two terms are singular as $P_{T}\to M(y)$. The first controls
the singular behavior near $y=0$ while the second modifies the singular
behavior for large $y$. The remaining terms constitute a polynomial
in $\log(P_{T})$ and $y^{2}$.
Thus, we parameterize 
the $y$-dependence with the set of functions $\{1, y^2\}$, and 
the $P_{T}$-dependence with the set of functions $\{1/L, 1, L \}$ 
where $L$ represents a logarithmic function of $P_{T}$. 
We believe that the  parameterization in terms of these 
$2\times 3=6$ functions 
is sufficient to reasonably describe the 
theoretical uncertainties.

Note that the coefficients of $f_{3}$ and $f_{4}$ are in the
ratio 10:1 and the coefficients of $f_{5}$ and $f_{6}$ are in the
ratio 5:1. While we could find an excellent fit without $f_{4}$ and
$f_{6}$, we retain these terms to provide flexibility when one tries
to fit the $\lambda_{J}$ coefficients to actual data.

We can perform a similar exercise for the LHC as well; these results
will be compiled and presented in the Section~\ref{sec:SummaryLHC}.

\section{Summation of Threshold Logs}

\begin{figure}[t]
\includegraphics[width=0.45\textwidth]{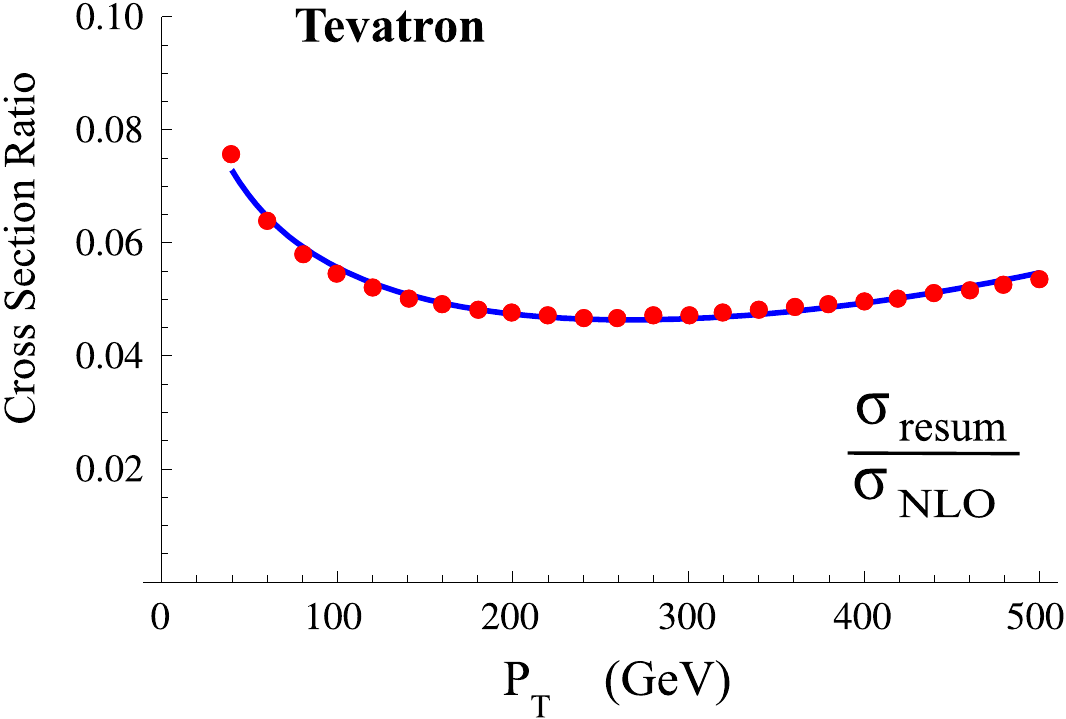}

\caption{The ratio of the two-loop threshold resummation contributions for
jet production compared to the total NLO cross section $\sigma_{\rm resum}/\sigma_{\rm NLO}$
at the Tevatron
($\sqrt{s}=1960$~GeV) vs. $P_T$ in GeV. 
We have set the scales to $\mu_{\rm F}=\mu_{\rm R}=P_{T}/2$,
and used $y=0$. The points are computed using the implementation
of the 2-loop threshold resummation by Kidonakis and Owens \citep{Kidonakis:2000gi}.\label{fig:resum}}

\end{figure}

For parton-parton scattering near the threshold for the production
of a jet with a given $P_{T}$, there is restricted phase space for
real gluon emission. Thus, there is an incomplete cancellation of
infrared divergences between real and virtual graphs, resulting in
large logarithms $L$ inside the integration over parton momentum fractions. 
At $n$-th order in $\alpha_{\rm s}$ these
logarithms enter the cross section in the general form $\alpha_{\rm s}^{n}\, L^{2n}$. The leading logarithms can be summed to all orders in $\alpha_{\rm s}$.
We make use of the numerical results from Ref.~\citep{Kidonakis:2001mc},
which has been implemented in the FastNLO program \citep{Kluge:2006xs}.

Fig.~\ref{fig:resum} displays the size of the threshold correction for
Tevatron jet measurements at $y = 0$. The curve is presented for the
scale choice $\mu=P_{T}/2$; we note that for this scale choice, the
threshold correction is generally smaller than with other scale
choices.\footnote{We do not present curves for $y = 1$ and $y = 2$
because these curves show a rise of the correction as $P_T$ decreases
from $200\ {\rm GeV}$, even though decreasing $P_T$ puts us farther
from the threshold. This rise is more pronounced for large $y$ than we
see for $y = 0$ in Fig.~\ref{fig:resum}. We suspect that this behavior
is an artifact of kinematic choices in the algorithm for summing
threshold logarithms, rather than being a real physical effect.}

We find the threshold corrections in this kinematic regime to be less
than those discussed in the previous section (Sec.~\ref{sec:ScaleDependence})
and shown in Figure~\ref{fig:muError0b}. As the threshold corrections
also arise from uncomputed higher-order terms, these corrections are,
in a sense, already accommodated by the larger uncertainty that we estimated
from scale variation in Eq.~(\ref{eq:errorSummary}). Indeed, the functions
$f_J$ for $J=1$ and $J=2$ contain singularities for $P_T \to M(y)$ that are meant to incorporate the threshold singularities. For this reason, we
will not add a separate $f_{J}(P_{T},y)$ function in the expression
for the total uncertainty ${\cal E}$ to represent the effects of threshold logarithms.

\section{Underlying event and hadronization
\label{sec:UEandHC}}

A separate source of uncertainties in jet measurements comes from
what is colloquially known as {}``splash-in'' and {}``splash-out''
corrections. {}``Splash-in'' corrections arise from the underlying
event, which can deposit additional energy into the jet cone; we will
refer to these more formally as underlying event (UE) corrections.
{}``Splash-out'' corrections come from the hadronization process
of the jet which may move some of the jet energy outside the defined
jet cone. We will refer to these as hadronization corrections (HC). 

In either case, the correction is modeled as adding an amount $\delta P_{T}$
to the observed transverse momentum (or transverse energy) of the
jet.  
We denote the average over many events of $\delta P_{T}$ by $\langle\delta P_{T}\rangle$. 
A complete analysis of the UE and HC contributions was performed by 
Cacciari, Dasgupta, Magnea, 
Salam in Refs.~\citep{Dasgupta:2007wa,Cacciari:2007aq,Dasgupta:2008di}. 
We find this to be an entirely suitable method 
for our estimate of $\langle\delta P_{T}\rangle$, and we 
adapt their results in the following. 

\subsection{Underlying event (UE) }

We can parameterize the effect of the underlying event corrections on the apparent $P_{T}$ of the jet as
\begin{equation}
\label{eq:ue}
\langle\delta P_{T}\rangle_{\rm UE}= 
\Lambda_{\rm UE}\,\frac{1}{2}\, R^{2} \quad ,
\end{equation}
where {\it R} is the cone radius of the jet and $\Lambda_{\rm UE}$ is the average transverse energy per unit rapidity in the underlying event. Because we model the {}``splash-in''
energy as random and uncorrelated with how the jet develops, the contribution
from the underlying event will scale as the area of the jet cone---hence
the factor of $R^{2}$ in Eq.~(\ref{eq:ue}). At Tevatron energies,
Ref.~\citep{Dasgupta:2007wa} finds 
\begin{equation}
\Lambda_{\rm UE}({\rm 1960\ GeV})\approx 3\pm1\ {\rm GeV}
\quad.\label{eq:lamUE}
\end{equation}
Thus, the $\left\langle P_{T}\right\rangle $ shift from the underlying
event corrections is given by
\begin{equation}
\left\langle \delta P_{T}\right\rangle _{\rm UE}\approx +0.7\,{\rm GeV}\pm 0.3\,{\rm GeV}\quad,
\label{eq:ptUE}
\end{equation}
for a jet cone with $R=0.7$.

\subsection{Hadronization correction (HC) }

The $R$ dependence of hadronization correction is very different from that of the underlying event correction \citep{Dasgupta:2007wa,Cacciari:2007aq,Dasgupta:2008di}. The smaller the jet cone is, the more likely it is that hadronization will spray hadrons out of the cone. Hence, we will parameterize these corrections as proportional to $1/R$. Following Ref.~\citep{Dasgupta:2007wa}, we write the
hadronization correction as 
\begin{equation}
\langle\delta P_{T}^{i}\rangle_{\rm HC}
=
-C_{i}\,\frac{2}{R}\,{\cal A(\mu_{I})}\label{eq:hc} \quad ,
\end{equation}
where ${\cal A}(\mu_{I})$ parameterizes the soft gluon radiation.
Ref.~\citep{Dasgupta:2007wa} takes $\mu_{I}=2\ {\rm GeV}$,
and finds ${\cal A}(2\ {\rm GeV})\approx 0.2\ {\rm GeV}$. In Eq.~(\ref{eq:hc}),
$C_{i}$ is a color factor that depends on whether the jet is initiated
by a quark, for which $C_i = C_{\rm F} = 4/3$, or by a gluon, for which $C_i = C_{\rm A} = 3$. We thus need an estimate
of the fraction of jets that are gluon jets. Using calculations from the literature \citep{Lai:1996mg}, 
we estimate that, for the Tevatron in the low $P_{T}$ region, the
fractions of quark and gluon jets are approximately
\begin{equation*}
f_{q} \approx  \frac{2}{3}\quad ,\hskip 1 cm
f_{g} \approx  \frac{1}{3}\quad.
\end{equation*}
Using these fractions, we can form a weighted average of the quark
and gluon terms to obtain 
\begin{eqnarray}
\langle\delta P_{T}\rangle_{\rm HC} & = & f_{q}\langle\delta P_{T}^{q}\rangle_{\rm HC}+f_{g}\langle\delta P_{T}^{g}\rangle_{\rm HC}
\nonumber \\
 & = & -f_{q}\frac{2C_{\rm F}}{R}{\cal A}(\mu_{I})
 -f_{g}\frac{2C_{\rm A}}{R}{\cal A}(\mu_{I})
\label{eq:ptHC}\\
& \approx  & -1\,{\rm GeV}\pm0.5\,{\rm GeV}
\quad.\nonumber 
\end{eqnarray}
 Here, we have used a typical cone radius of $R=0.7$ and taken a
conservative choice for the uncertainty of 50\% of the correction.

\subsection{$\langle\delta P_{T}\rangle$ from the UE and HC }

\begin{figure}[t]
\includegraphics[width=0.45\textwidth]{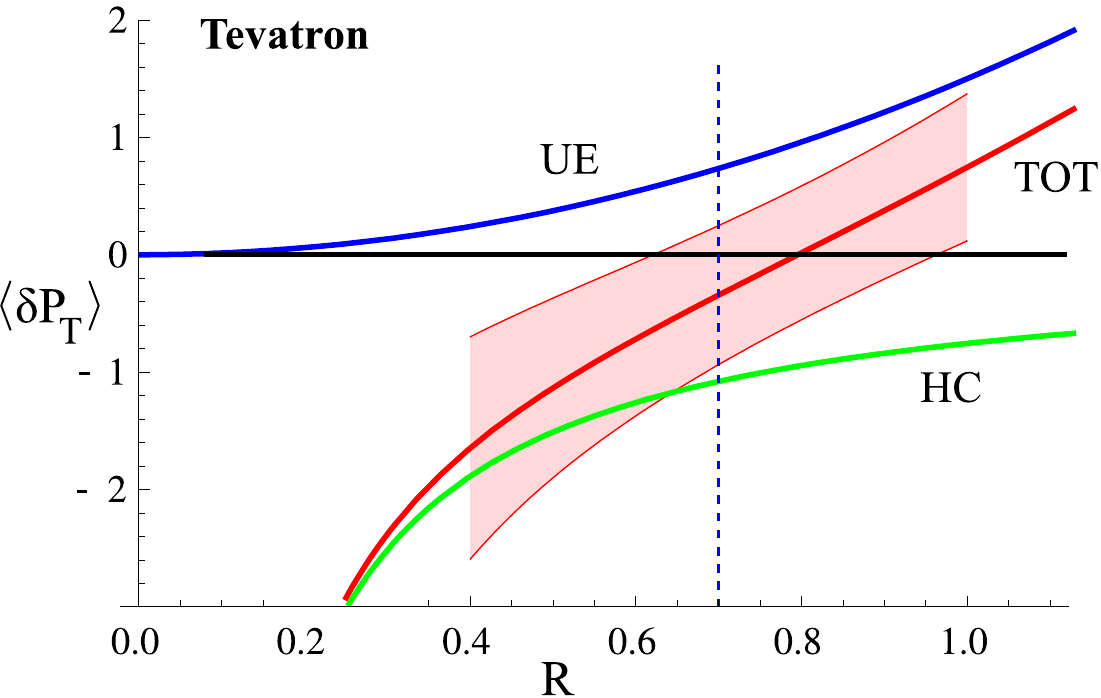}
\caption{We display the expected $P_{T}$ shift, $\left\langle \delta P_{T}\right\rangle $,
in GeV vs. jet cone radius $R$ for the UE, HC, and combined results (TOT)
at the Tevatron. The calculation of the HC uses a combination of quark-initiated
($f_{q}=2/3$) and gluon-initiated ($f_{g}=1/3$) jets. The upper
solid (blue) line represents the UE correction, and the lower solid
(green) line represents the HC terms. The combination of these corrections (TOT)
is represented by the central (red) band including the uncertainties.  
The vertical line corresponds to $R$=0.7.
}
\label{fig:uehcTEV} 
\end{figure}

Combining the underlying event of Eq.~(\ref{eq:ptUE}) and the hadronization
corrections of Eq.~(\ref{eq:ptHC}), the net $P_{T}$ shift is \begin{equation}
\langle\delta P_{T}\rangle\approx  -0.3\,{\rm GeV}\pm 0.6\,{\rm GeV}\;\;,\label{eq:deltaET}\end{equation}
 where we have added the separate uncertainties in quadrature.

The individual underlying event and hadronization
results for $\langle\delta P_{T}\rangle$
are displayed in Fig.~\ref{fig:uehcTEV} for the Tevatron using the
parameterizations of Eq.~(\ref{eq:ptUE}) and Eq.~(\ref{eq:ptHC}).
The combined result for $\langle\delta P_{T}\rangle$, including the
uncertainty band, is also displayed. The underlying event and hadronization
corrections have opposite sign, and we note that for a jet cone radius
of $R = 0.7$, the two corrections nearly cancel each
other.

\subsection{From $\langle\delta P_{T}\rangle$ to $\delta\sigma$}

\begin{figure}[t]
\includegraphics[width=0.45\textwidth]{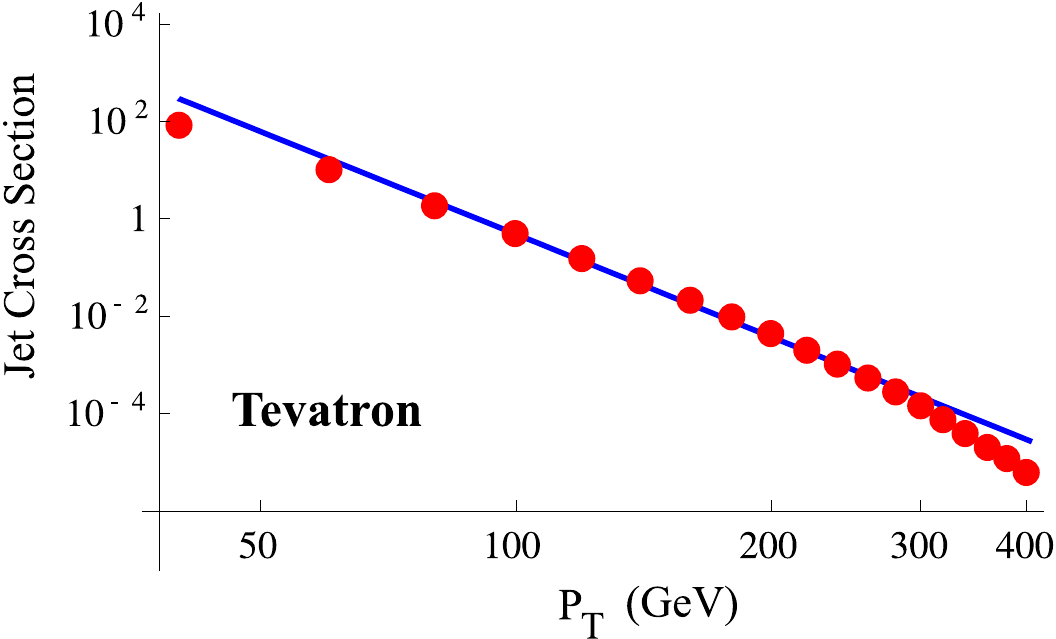}
\caption{Jet cross section $d^{2}\sigma/dP_{T}/dy$ vs. $P_{T}$ in GeV with $y=0$
at the Tevatron  in units of ${\rm nb/GeV}$. The line is a power law fit
with $n=7$; this describes the slope of the jet data in the 
range $P_{T}\approx [50,\,300]$~GeV. \label{fig:jetCrossSectionPower}}
\end{figure}

The differential jet cross section can be  approximated by a  
power law of the form

\begin{equation}
\frac{d\sigma(P_{T})}{dP_{T}}\approx\frac{{\it const.}}{P_{T}^{n}}
\quad.
\label{eq:powerlaw}
\end{equation}
in the specific $P_{T}$ range of interest.
For jets at the Tevatron 
in the intermediate $P_{T}$ range of $\sim[50,\,300]$~GeV,
we find $n\approx7$ as illustrated by Fig.~\ref{fig:jetCrossSectionPower}.

The effect of the underlying event and hadronization corrections is to shift
the jet $P_{T}$ from its value $P_{T}^{\rm pert}$ at the NLO parton level to a new value
\[
P_{T}=P_{T}^{{\rm pert}} + \langle\delta P_{T}\rangle
\quad,
\]
where $\langle\delta P_{T}\rangle$ is the average change in the transverse jet transverse momentum due to underlying event additions and hadronization subtractions from Eq.~(\ref{eq:deltaET}).

If we write the true differential cross section as a function $f$,
\[
\frac{d\sigma(P_{T})}{dP_{T}}\equiv f(P_{T})
\quad,
\]
then $f$ is related to the perturbatively calculated function $f_{{\rm pert}}$ by
\[
f(P_{T})\approx f_{{\rm pert}}(P_{T}^{{\rm pert}})
=f_{{\rm pert}}\big(P_{T} - \langle\delta P_{T}\rangle \big)
\quad.
\]
We can perform a Taylor expansion about $P_{T}$ for small $\delta P_{T}$,
\begin{eqnarray*}
f(P_{T}) & \approx & 
f_{{\rm pert}} \big(P_{T}-\langle\delta P_{T}\rangle \big)
\\
 & \approx  & f_{{\rm pert}}(P_{T})
 -\langle\delta P_{T}\rangle \,\frac{df_{{\rm pert}}'(P_{T})}{dP_{T}}
\\
 & = & f_{{\rm pert}}(P_{T})\left\{ 1+n\,\frac{\langle\delta P_{T}\rangle}{P_{T}}\right\} 
\quad.
\end{eqnarray*}
Here we have used the power law of Eq.~(\ref{eq:powerlaw}) to replace
$f'(P_{T})$ by $-n\, f(P_{T})/P_{T}$. Thus, to first order we find%
\footnote{\emph{Cf.}, Eq.~(5.9) of Dasgupta \emph{et al.} in Ref.~\citep{Dasgupta:2007wa}} 
\begin{equation}
\frac{d\sigma}{dP_{T}}\approx \frac{d\sigma_{{\rm pert}}}{dP_{T}}\left[1+n\,\frac{\langle\delta P_{T}\rangle}{P_{T}}+\cdots\right]
\;\;,
\label{eq:ptSIG}\end{equation}
so that the fractional correction is $n\,\langle\delta P_{T}\rangle/P_{T}$.
Using $n\approx7$ and the estimate from Eq.~(\ref{eq:deltaET}) of $\langle\delta P_{T}\rangle$, we find that the fractional correction to the cross section is approximately
\begin{equation*}
7\ \times \ 
\frac{- 0.3\,{\rm GeV} \pm 0.6\,{\rm GeV}}{P_T}
\approx
-\frac{2\,{\rm GeV}}{P_T} \pm 
\frac{4\,{\rm GeV}}{P_T}
\;\;.
\end{equation*}
Thus we estimate the fractional uncertainty from the underlying event and hadronization to be ${4\,{\rm GeV}}/{P_T}$.

We account for this source of uncertainty by adding a new function $f_J(P_T,y)$ with $J=7$,
\begin{equation}
f_{7}(P_{T},y)=\frac{4\ {\rm GeV}}{P_{T}}
%\quad.
%
\end{equation}
for Tevatron jets in the  $P_{T}$ range of $\sim[50,\,300]$~GeV.

\section{Summary for the Tevatron\label{sec:SummaryTEV}}

\begin{figure*}[t]
\includegraphics[clip,width=0.32\textwidth]{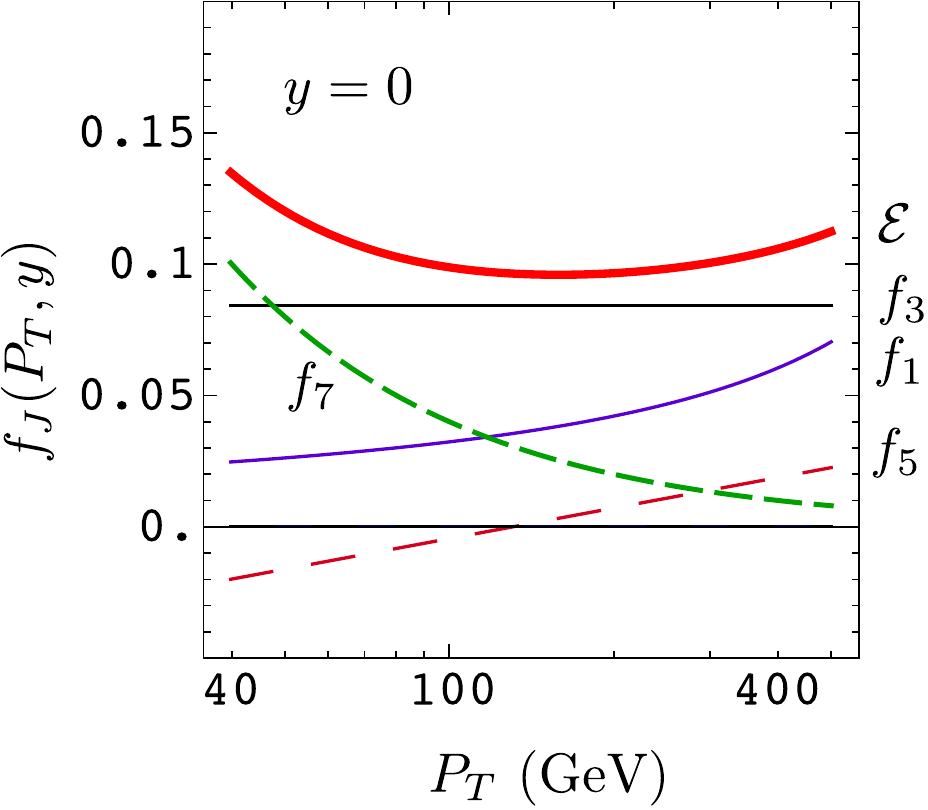}\hspace{0.01\textwidth}
\includegraphics[clip,width=0.32\textwidth]{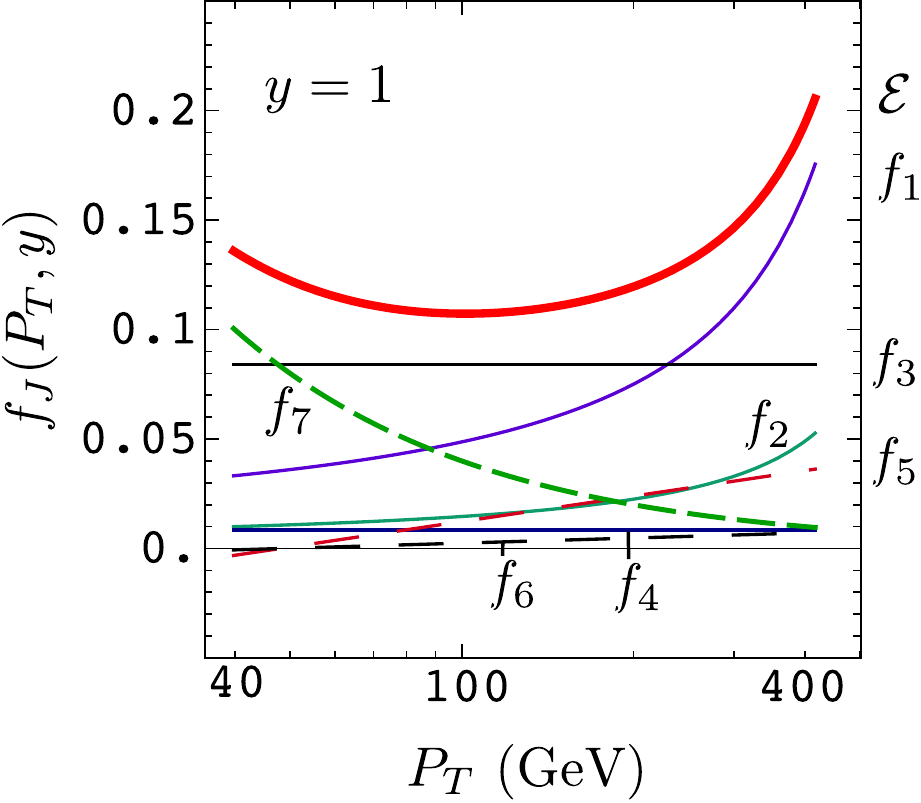}\hspace{0.01\textwidth}
\includegraphics[clip,width=0.32\textwidth]{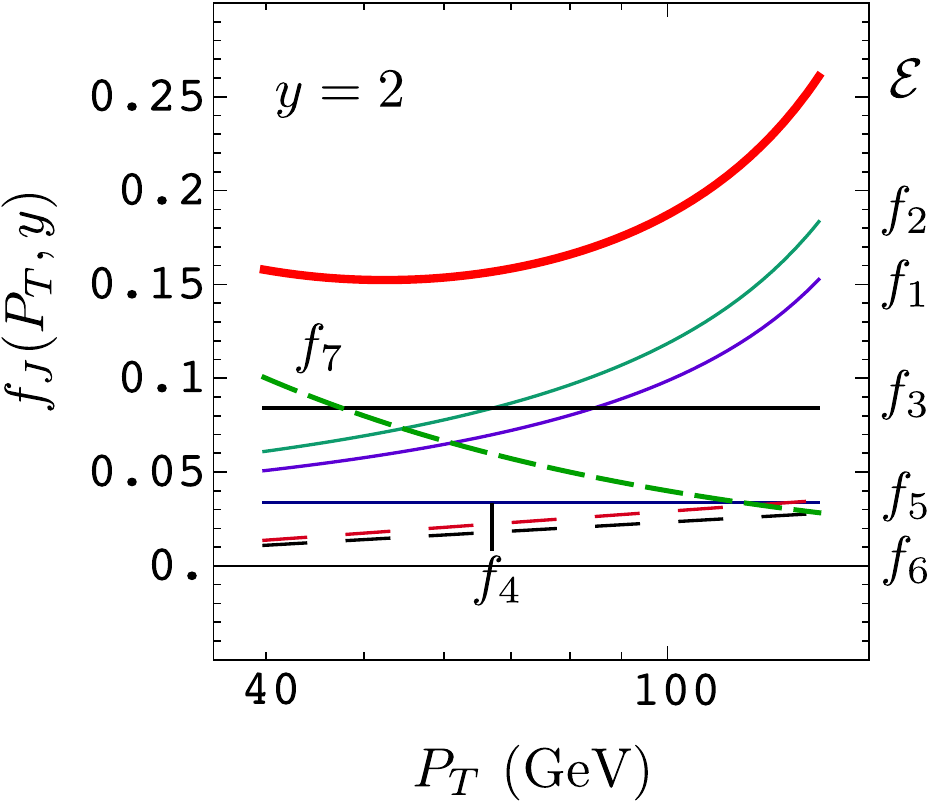}
\caption{A compilation of the uncertainties for jet production at the the Tevatron
($\sqrt{s}=1960$~GeV) for $y=\{0,1,2\}$. The numeric label corresponds
to the error components summarized in Eq.~(\ref{eq:errorSummaryLHC}).
The upper thick (red) line is the quadrature sum of the individual
errors .\label{fig:summaryTEV}}

\end{figure*}

\begin{table}[!t]
\begin{tabular}{|c|c|}
\hline 
Uncertainty $f_J$  & Source\tabularnewline
\hline
\hline 
$\{f_{1},f_{2},f_{3},f_{4},f_{5},f_{6}\}$  &  perturbative \tabularnewline
\hline 
$f_{7}$  & non-perturbative \tabularnewline
\hline
\end{tabular}

\caption{A compilation of the source of uncertainties ($f_J$) that
comprise the total jet cross section uncertainty ${\cal E}$. 
The perturbative uncertainties arise from the higher, uncalculated,
orders of perturbation theory and are estimated using the
$\{\mu_{\rm F},\mu_{\rm R}\}$ scale variation of the calculated cross
section.
The non-perturbative uncertainties are an estimate of the
underlying event and hadronization corrections.
\label{tab:f}
}
\end{table}

We have described the correlated theoretical systematic uncertainty
using a total of seven functions, as summarized in Table~\ref{tab:f}.
The net error at any one value of $\{P_{T},y\}$ is obtained by adding
these seven functions in quadrature \begin{equation}
{\cal E}(P_{T},y)\equiv\sqrt{\sum f_{J}(P_{T},y)^{2}}\;\;.\label{eq:totalerror}\end{equation}

We now summarize the complete set of contributions to the uncertainty
of the differential jet cross section as a function of $\{P_{T},y\}$
for the Tevatron:
\begin{equation}
\begin{split}f_{1}(P_{T},y)={} & \frac{9.62\times10^{-2}}{\log(M(y)/P_{T})}\;\;,\\
f_{2}(P_{T},y)={} & \frac{2.89\times10^{-2}\,\, y^{2}}{\log(M(y)/P_{T})}\;\;,\\
f_{3}(P_{T},y)={} & 8.42\times10^{-2}\;\;,\\
f_{4}(P_{T},y)={} & 0.842\times10^{-2}\, y^{2}\;\;,\\
f_{5}(P_{T},y)={} & 1.68\times10^{-2}\,\,\log\!\left(\frac{15\, P_{T}}{M(y)}\right)\;\;,\\
f_{6}(P_{T},y)={} & 0.336\times10^{-2}\,\, y^{2}\log\!\left(\frac{15\, P_{T}}{M(y)}\right)\;\;,\\
f_{7}(P_{T},y)={} & \frac{4\, {\rm GeV}}{P_{T}}\;\;.\end{split}
\label{eq:errorSummaryTEV}\end{equation}
We display these results in Figure~\ref{fig:summaryTEV}. For $P_{T}\gtrsim100\ {\rm GeV}$,
the perturbative uncertainties are dominant, and slowly
rise with increasing $P_{T}$; this results holds across the full
$y$-range, but the rise with $P_T$ is more pronounced at large $y$. For $P_{T}\lesssim100\ {\rm GeV}$, the uncertainty
from the UE and HC terms become increasingly important as $P_{T}$ decreases.

\section{Theory errors at the LHC\label{sec:SummaryLHC}}

Having demonstrated the method for determining the theoretical systematic
uncertainty at the Tevatron, we perform a parallel analysis for the
Large Hadron Collider (LHC).

\subsection{Perturbative uncertainty}

We again estimate the error from not having calculated beyond NLO by using the dependence of the NLO cross section on the scales $\{\mu_{{\rm R}},\mu_{{\rm F}}\}$, just as in the Tevatron case, and this yields the functions
$\{f_{1},f_{2},f_{3},f_{4},f_{5},f_{6}\}$ summarized in Eq.~(\ref{eq:errorSummaryLHC}) at the end of this section.

% \subsection{Summation of Threshold Logs}
% 
% As for the Tevatron case, we have used the threshold corrections as
% represented by the program of Ref.~\citep{Kidonakis:2001mc} as an
% independent estimate of the size of perturbative contributions to the
% cross section that are beyond next-to-leading order. We find for $y=0$ that this
% estimate is consistent with the estimate derived from the scale
% variation of the NLO result, represented in the functions
% $\{f_1,\dots, f_6\}$ in Eq.~(\ref{eq:errorSummaryLHC}).
% 

\subsection{Underlying event and hadronization}

We proceed as in Sec.~\ref{sec:UEandHC} for the Tevatron, accounting for
the changed circumstances at the LHC. We first need to estimate the error in the determination of the contribution to the average jet transverse momentum, $\langle \delta P_{T}\rangle$, arising from the underlying event and from hadronization. 

The underlying event contribution to $\langle \delta P_{T}\rangle$ is determined by the parameter $\Lambda_{\rm UE}$ in Eq.~(\ref{eq:ue}). 
Consistently with Refs.~\citep{Dasgupta:2007wa,Cacciari:2007aq,Dasgupta:2008di},
for the LHC we take $\Lambda_{\rm UE}({\rm 14\ TeV})\approx 10\pm 4$~GeV,
and obtain
\begin{equation}
\left\langle \delta P_{T}\right\rangle _{\rm UE}\approx +2.5\,{\rm GeV}\pm1\,{\rm GeV}
\quad.
\label{eq:UEptLHC}
\end{equation}

For the contribution to $\langle \delta P_{T}\rangle$ from hadronization, we use Eq.~(\ref{eq:ptHC}) with ${\cal A}(\mu_I)\approx  0.2\ {\rm GeV}$ as before. For the fractions $f_{q}$ and $f_{g}$ of quark and gluon jets in the relatively low $P_T$ region where the hadronization corrections are significant, we use
\begin{equation*}
f_{q} \approx  \frac{1}{3}\quad ,\hskip 1 cm
f_{g} \approx  \frac{2}{3}\quad.
\end{equation*}
Using these fractions, we can form a weighted average of the quark
and gluon terms and estimate the hadronization contribution to $\langle \delta P_{T}\rangle$ to be
\begin{equation}
\left\langle \delta P_{T}\right\rangle _{\rm HC}   =   -1.4\,{\rm GeV}\pm0.7\,{\rm GeV}
\quad.
\label{eq:HCptLHC}
\end{equation}

Combining the underlying event and hadronization contributions, we
estimate
\begin{equation}
\left\langle \delta P_{T}\right\rangle \approx +1\,{\rm GeV}\pm 1.2\,{\rm GeV}
\;\;,
\label{eq:ptLHC}
\end{equation}
where we have added the separate uncertainties in quadrature.

\begin{figure}[t]
\includegraphics[width=0.45\textwidth,keepaspectratio]{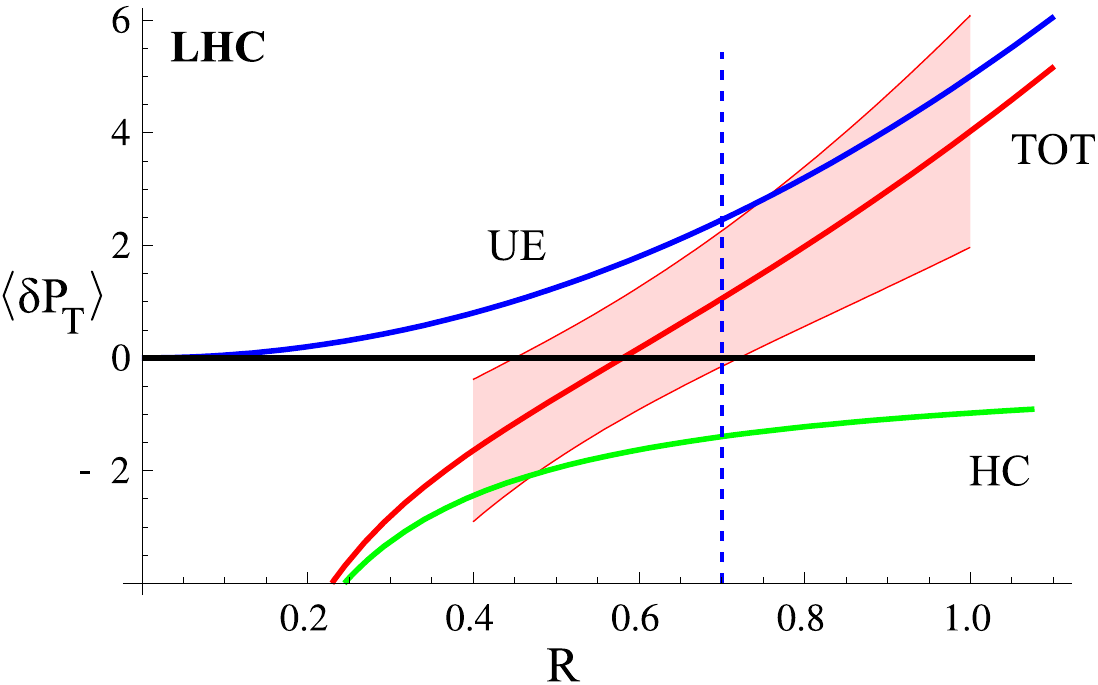}

\caption{We display the expected $P_{T}$ shift, $\left\langle \delta P_{T}\right\rangle $,
in GeV vs. jet cone radius $R$ for the UE, HC, and combined results (TOT)
at the LHC. The calculation of the HC uses a combination of quark-initiated
($f_{q}=1/3$) and gluon-initiated ($f_{g}=2/3$) jets. The upper
solid (blue) line represents the UE correction, and the lower solid
(green) line represents the HC terms. The combination of these corrections (TOT)
is represented by the central (red) band including the uncertainties. 
The vertical line corresponds to $R$=0.7.
}

\label{fig:uehcLHC} 
\end{figure}

The results for the underlying event and hadronization contribution to 
$\langle \delta P_{T}\rangle$ are displayed in Fig.~\ref{fig:uehcLHC} for the LHC using the parameterizations of Eq.~(\ref{eq:HCptLHC}) and Eq.~(\ref{eq:UEptLHC}) but with a variable cone size $R$.

\begin{figure}[t]
\includegraphics[width=0.45\textwidth]{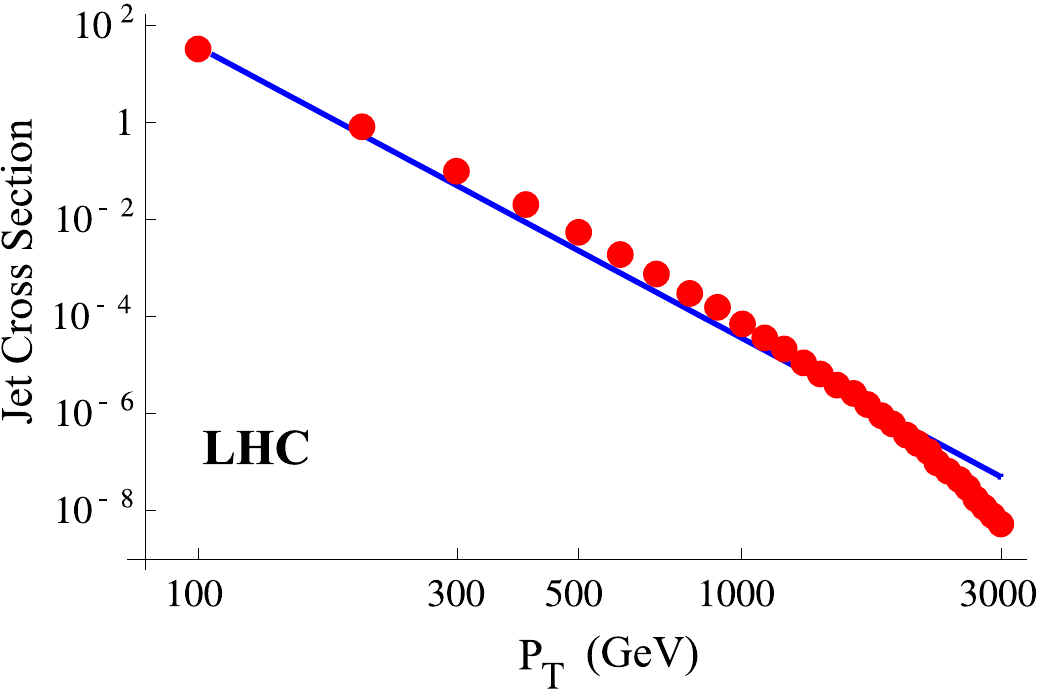}

\caption{Jet cross section $d^{2}\sigma/dP_{T}/dy$ vs. $P_{T}$ in GeV with $y=0$
at the LHC ($\sqrt{s}=14$~TeV) in units of ${\rm nb/GeV}$. The line is
a power law fit with $n=6$; this describes the slope of the jet data
in the  range $P_{T}\approx [100,\,1000]$~GeV. \label{fig:lhcJetPower}}

\end{figure}

The correction to $\langle \delta P_{T}\rangle$ determines the correction to the cross section via Eq.~(\ref{eq:ptSIG}). For this, we need the power $n$ that describes the approximate power law fall-off of the cross section. As illustrated in Fig.~\ref{fig:lhcJetPower}, a power law with $n\approx 6$
describes the data over the range $P_{T} \approx  [100, 1000] \ {\rm GeV}$.
Using $n\approx6$ and the estimate from Eq.~(\ref{eq:ptLHC}) of $\langle\delta P_{T}\rangle$, we find that the fractional correction to the cross section is approximately
\begin{equation*}
6\ \times \ 
\frac{1\,{\rm GeV} \pm 1.2\,{\rm GeV}}{P_T}
\approx
\frac{6\,{\rm GeV}}{P_T} \pm 
\frac{7\,{\rm GeV}}{P_T}
\;\;.
\end{equation*}
Thus we estimate the fractional uncertainty from the underlying event and hadronization to be ${7\,{\rm GeV}}/{P_T}$. 
We include this in the estimate of systematic theoretical errors by including a function $f_7(P_T)$ given by
\begin{equation}
f_{7}(P_{T})=\frac{7\ {\rm GeV}}{P_{T}}
%\quad.
%
\end{equation}
for LHC jets in  the range $P_{T} \approx  [100, 1000] \ {\rm GeV}$.

\subsection{Summary: LHC }

\begin{figure*}[t]
\includegraphics[width=0.30\textwidth]{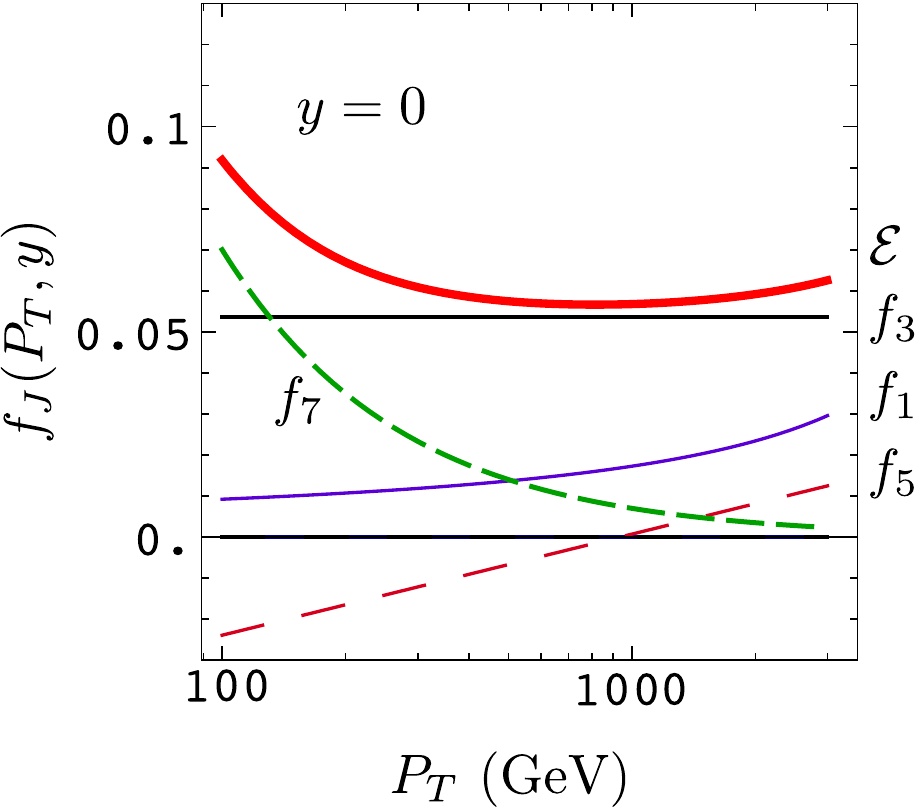} \hskip 0.03\textwidth
\includegraphics[width=0.31\textwidth]{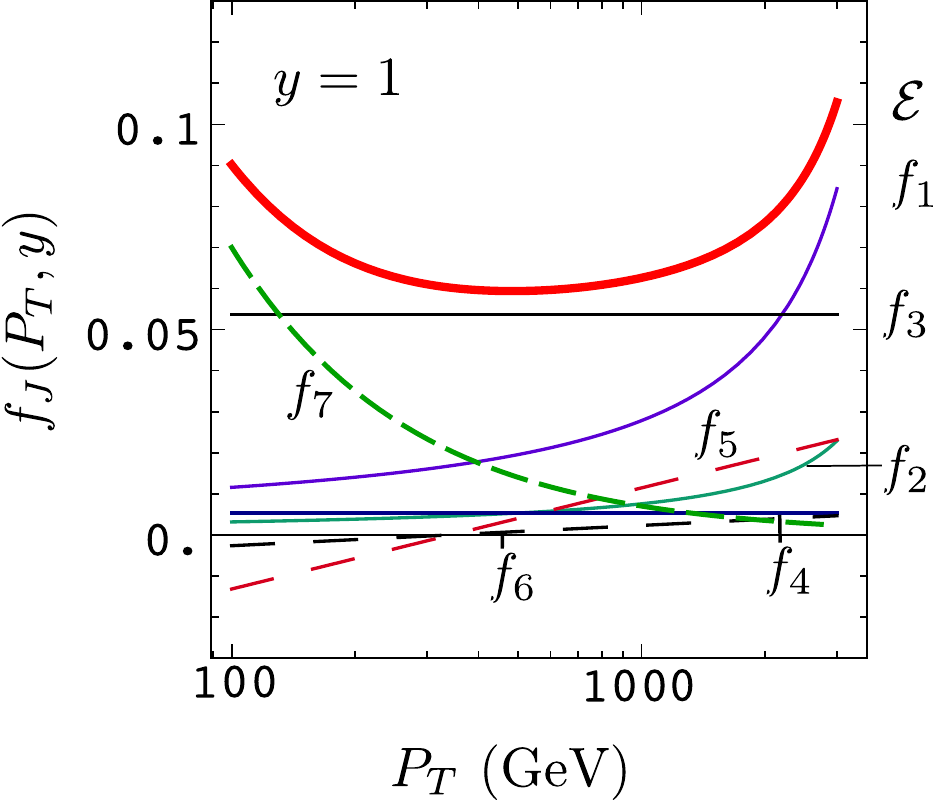} \hskip 0.03\textwidth
\includegraphics[width=0.30\textwidth]{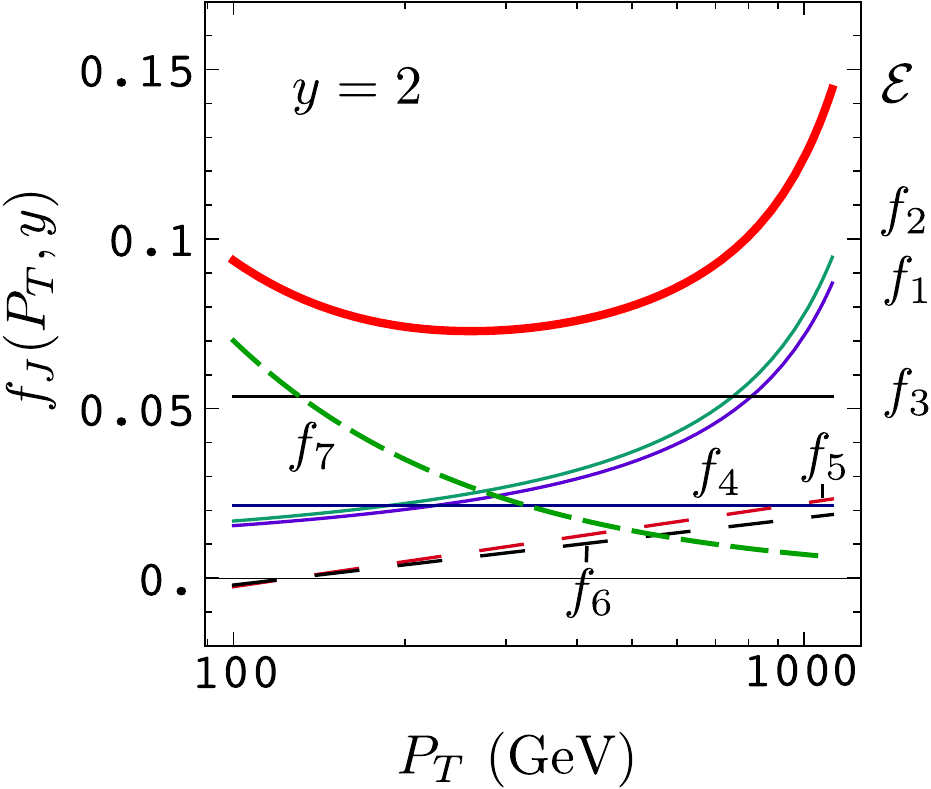}
\caption{A compilation of the uncertainties for jet production at the LHC ($\sqrt{s}=14,000$~GeV)
for $y=\{0,1,2\}$. The numeric label corresponds to the error components
summarized in Eq.~(\ref{eq:errorSummaryLHC}). The upper thick (red)
line is the quadrature sum of the individual errors. \label{fig:summaryLHC}}
\end{figure*}

We now summarize the complete set of contributions to the uncertainty
of the differential jet cross section as a function of $\{P_{T},y\}$
for the LHC: 
\begin{equation}
\begin{split}
f_{1}(P_{T},y)={} & \frac{4.56\times10^{-2}}{\log(M(y)/P_{T})}\;\;,\\
f_{2}(P_{T},y)={} & \frac{1.24\times10^{-2}\, y^{2}}{\log(M(y)/P_{T})}\;\;,\\
f_{3}(P_{T},y)={} & 5.36\times10^{-2}\;\;,\\
f_{4}(P_{T},y)={} & 0.536\times10^{-2}\, y^{2}\;\;,\\
f_{5}(P_{T},y)={} & 1.07\times10^{-2}\,\log\!\left(\frac{15\, P_{T}}{M(y)}\right)\;\;,\\
f_{6}(P_{T},y)={} & 0.214\times10^{-2}\, y^{2}\log\!\left(\frac{15\, P_{T}}{M(y)}\right)\;\;,\\
f_{7}(P_{T},y)={} & \frac{7\, {\rm GeV}}{P_{T}}
\;\;.
\end{split}
\label{eq:errorSummaryLHC}
\end{equation}

We display these results in Figure~\ref{fig:summaryLHC}. In the
central rapidity ($y\sim0$) region for $P_{T}\gtrsim500\ {\rm GeV}$
the perturbative uncertainties are dominant and slowly
rise with increasing $P_{T}$, while for $P_{T}\lesssim 500\ {\rm GeV}$
the nonperturbative uncertainties become increasingly important. For $y=2$, the transition $P_{T}$ is closer
to 300~GeV than 500~GeV.

\section{Conclusions}

As the LHC prepares to take data, it is important
that we be able to determine whether a physics signal is consistent
with the standard model. For example, if we observe a signal
that is inconsistent with the standard model prediction, but the augment for this inconsistency includes \emph{only} experimental
errors, we cannot claim this is {}``new physics'' until we demonstrate
it is also inconsistent including \emph{both} experimental
and theoretical errors. This paper provides a framework to quantitatively
make such a determination in the case of jet physics. Similarly, this paper provides a framework to quantitatively fit parton distribution functions to Tevatron and LHC jet data, including estimated errors from the theory.

The framework that we provide involves functions $f_{J}(P_{T},y)$ that represent independent contributions to the theory error. We note that other authors might estimate the errors differently and thus produce different functions $f_{J}(P_{T},y)$. We hope that this will happen and that the merit of different choices will be debated.

\goodbreak
\begin{acknowledgments}
We thank Z.~Nagy, M.~Dasgupta, L.~Magnea, S.~Mrenna, P.~Nadolsky,
and J.~F.~Owens for valuable discussions. We acknowledge the hospitality
of CERN and LPSC Grenoble 
where a portion of this work was performed. This work is supported
by the U.S. Department of Energy under grants DE-FG02-04ER41299, DE-FG02-96ER40969,
and the Lightner-Sams Foundation.
\end{acknowledgments}

\bibliographystyle{hunsrt}
\bibliography{bibjet}

\end{document}